\documentclass[aps,superscriptaddress,showpacs,floatfix,twocolumn]{revtex4-1}
\usepackage{amsmath,amssymb,amsfonts,bbm,graphicx,color,times}
\usepackage{subfigure}
\usepackage[english]{babel}
\usepackage{float}
\usepackage{color,soul}
\newcommand{\bra}[1]{\ensuremath{\langle #1|}}
\newcommand{\ket}[1]{\ensuremath{|#1 \rangle}}

\newcommand{\ketbra}[2]{\ensuremath{| #1 \rangle\hspace{-2pt} \langle #2 |}}

\newcommand{\eref}[1]{(\ref{#1})}

\newcommand{\Fref}[1]{Figure \ref{#1}}

\newcommand{\llrr}[1]{\ensuremath{\left( #1\right)}}
\newcommand{\llrrq}[1]{\ensuremath{\left[ #1\right]}}

\DeclareMathOperator{\Span}{span}

\begin{document}
\title{Characterization of qubit chains by Feynman probes}
\author{Dario Tamascelli}
\email{dario.tamascelli@unimi.it}
\affiliation{Quantum Technology Lab, Dipartimento di Fisica, 
Universit\`a degli Studi di Milano, I-20133 Milano, Italy}
\affiliation{Institut f\"ur Theoretische Physik \& IQST, Albert-Einstein-Allee 11, Universit\"at Ulm, Germany}
\author{Claudia Benedetti}
\email{claudia.benedetti@unimi.it}
\affiliation{Quantum Technology Lab, Dipartimento di Fisica, 
Universit\`a degli Studi di Milano, I-20133 Milano, Italy}
\author{Stefano Olivares}
\email{stefano.olivares@unimi.it}
\affiliation{Quantum Technology Lab, Dipartimento di Fisica, 
Universit\`a degli Studi di Milano, I-20133 Milano, Italy}
\affiliation{INFN, Sezione di Milano, I-20133 Milano, Italy}
\author{Matteo G. A. Paris}
\email{matteo.paris@fisica.unimi.it}
\homepage{http://users.unimi.it/aqm}
\affiliation{Quantum Technology Lab, Dipartimento di Fisica, 
Universit\`a degli Studi di Milano, I-20133 Milano, Italy}
\affiliation{INFN, Sezione di Milano, I-20133 Milano, Italy}
\date{\today}
\begin{abstract}
We address the characterization of qubit chains 
and assess the performances of local measurements 
compared to those provided by {\em Feynman probes}, 
i.e. nonlocal measurements realized by coupling a single 
qubit register to the chain. We show that local measurements 
are suitable to estimate small values of the coupling and that 
a Bayesian strategy may be successfully exploited to achieve 
optimal precision. For larger values of the coupling 
Bayesian local strategies do not lead to a consistent estimate.
In this regime, Feynman probes may be exploited to build 
a consistent Bayesian estimator that saturates the 
Cram\'er-Rao bound, thus providing an effective 
characterization of the chain.
Finally, we show that ultimate bounds to 
precision, i.e. saturation of the {\em quantum}
Cram\'er-Rao bound, 
may be achieved by a two-step scheme employing Feynman probes
followed by local measurements.
\end{abstract}
\pacs{03.67.-a, 75.10.Pq}
\maketitle
\section{Introduction}
Spin networks and strongly coupled systems of qubits are crucial
building blocks for large scale quantum computers \cite{lsq1,lsq2}. 
They also represent a
resource for short distance quantum communication \cite{sgb,kay10},
state transfer \cite{chr04,cpv07,cdf08,ya011} and quantum engineering,
e.g. generation of entanglement between distant qubits
\cite{cpv06,dam07,cub08,ban11,richerme14,saikat16}. These tasks usually require 
fine-tuning of the interaction parameters and, in turn, a precise
characterization of the spin coupling.  
Coupling constants, however, are often unaccessible in a direct
way, either because of experimental impediments or because they 
do not correspond to any proper observable.  This happens
for several quantities of interest in quantum technology and
in all these cases, quantum estimation theory \cite{books,Giov06,lqe4}
provides tools to evaluate the ultimate precision
attainable by any estimation procedure and to design optimal
measurement schemes.  Examples include the estimation
of the phase \cite{monras06,genoni11,delgado14,jacopo15}, quantum
correlations \cite{gen08,blandino12,benedetti13}, temperature
\cite{brunelli12, correa15}, characterization of classical processes or
environmental parameters \cite{benedettiG, benedettiNG, zwick15,zwick16}, 
and, indeed, the coupling constants of different kinds of interactions
\cite{carmen08, dariano01,geno12, stenberg14, bina16}.
\par
Here, we address the characterization of qubit systems made of linear
chains of coupled two-level systems, with emphasis on strongly coupled
ones, and assess performances of local measurements compared to {\em
Feynman probes}, i.e. nonlocal measurements realized by entangling a
single qubit register to the chain of qubits. 
The Feynman probes implement the idea of characterizing complex systems, 
with many degrees of freedom, by coupling
them to a simple quantum system, such as a qubit in our case, whose
dynamics depends on the features of the complex systems we want to
describe 
\cite{nok16,cosco16,rossi16,troiani16}.
By performing measurements on the quantum probe, we are able
to extract useful information about the system, causing minimal
disturbance.
\par
In this work we show that local measurements provide optimal
characterization for small values of the coupling constant, whereas for
larger couplings Feynman probes allow one to build 
a consistent Bayesian estimator that saturates the Cram\'er-Rao 
(CR)
bound, i.e. provides an effective characterization of 
the qubit system.  We also show that estimation 
by Feynman probes, complemented by local measurement,
represents an optimal characterization scheme for strongly coupled qubit
systems, achieving the ultimate bound to precision. 
Indeed, nonlocal measurements have already been  suggested as a
convenient toolbox for quantum circuits based on trapped ions
\cite{ti1,ti2} and 
superconducting qubits \cite{nig13,blu16}.
\par
The system we are going to investigate is a linear lattice of equally 
coupled two-level systems $\vec{\sigma}^j = \llrr{\sigma_x^j,
\sigma_y^j,\sigma_z^j}$,  where $\sigma_k^j$ denotes the Pauli matrix in direction $k=x,y,z$ for the $j^{th}$ particle  and $j=1,2,\ldots,s$, whose interaction 
Hamiltonian is given by
\begin{equation}
    {\cal H}_0 = -\frac{\nu}{2} \sum_{j=1}^{s-1} 
    \sigma_+^{j+1} \sigma_-^j  +  \sigma_+^{j} \sigma_-^{j+1} 
    \label{eq:hamIntFull}
\end{equation}
where $ \sigma_\pm^j = \frac12 (\sigma_x^j \pm i \sigma_y^j)$ and $\nu$ is the
coupling constant between nearest neighbor spins. 
The Hamiltonian ${\cal H}_0$ preserves the number $N_z = \sum_{j=1}^s 
\frac12 (\mathbb{I}+ \sigma_z^j)$ of ``up'' spins, i.e.
\begin{equation}
    [{\cal H}_0,N_z]=0.
    \label{eq:conservation}
\end{equation}
The characterization of the system amounts to the determination
of the unknown value of the effective coupling $\lambda=\nu\tau$,  with
$\tau$ being the interaction time from the initialization of the 
chain. To this aim, we focus 
on initial preparations of the system where a single spin is up,
whereas all the other are down. We will refer to the single spin up 
as  the \emph{excitation} of the chain. Thanks to the 
conservation law \eref{eq:conservation}, the Hamiltonian 
\eref{eq:hamIntFull}, restricted to the single-excitation subspace, can be rewritten as 
\begin{equation}
     {\cal H}_0 = -\frac{\nu}{2} \sum_{j=1}^{s-1} 
     \ketbra{j+1}{j} + \ketbra{j}{j+1}\,,
    \label{eq:H0}
\end{equation}
where $|j\rangle$ denotes a state having an excitation at site $j$ and 
the set $\{\ket{j}\}$ constitutes an orthonormal basis in the single-excitation subspace.
The eigenvalues and eigenvectors of  the Hamiltonian \eref{eq:H0} are:
\begin{align}
    &e_k(\nu) = -\nu \cos\llrr{\frac{k \pi}{s+1}}, \label{eq:freeVal} \\
   &\ket{e_k} =  \sqrt{\frac{2}{s+1}}\sum_{j=1}^s \sin\left (\frac{k
   \pi j}{s+1} \right) \ket{j}. \label{eq:freeSpec}
\end{align}
In the following we analyze and compare different strategies  
for the estimation of the effective coupling parameter $\lambda$  and
also assess their precision against the
ultimate bounds posed by quantum mechanics itself.
\par
This paper is structured as follows. In Sec. \ref{optm} we review the main tools
of the quantum estimation theory; in Sec. \ref{locmesS}, we apply these tools to the 
estimation of the coupling constant of the chain by a local measurement on a single site of the chain. In Sec. \ref{FePS}, we
introduce the concept of Feynman probes  and we evaluate the associated Fisher information.
In Sec. \ref{bayesS}, we present the results of a simulated set of repeated mesurements on the system, both local and using a Feynman probe, to estimate the coupling constant of the qubit lattice 
and compare their performances by evaluating their variances. Section \ref{conlS} closes the paper with  final remarks and  discussion.
\section{Optimal measurement}\label{optm}
 The performances of an estimation procedure in terms of precision may
be assessed by the Fisher information of the associated distribution.
The Fisher information on the parameter $\lambda$ carried by an
observable random variable whose distribution depends on the parameter
$\lambda$, i.e. $X\sim p(x|\lambda)$, is defined as:
\begin{equation}
    F(\lambda) = E\llrrq{\llrr{
    \frac{\partial}{\partial \lambda}\ln p(x|\lambda)}^2}\,,
    \label{eq:defFisher}
\end{equation}
where $E(...)$ denotes the expectation value over the distribution $p(x|\lambda)$.
\par
The Fisher information sets the lower bound for the variance of any unbiased 
estimator $\lambda(x_1, x_2,...)$ of the parameter $\lambda$, based on the outcomes of
$X$ through the CR inequality:
\begin{equation}
\hbox{Var} \lambda \geq \frac{1}{M F(\lambda)}\,,
    \label{eq:cramerrao}
\end{equation}
where $M$ is the number of repeated measurements.
Estimators saturating the CR inequality are referred to as \emph{efficient} estimators.
\par
In a quantum setting, a measurable quantity corresponds to an observable
$A=\sum a\ketbra{a}{a}$ on some Hilbert space $\mathcal{H}$, whose statistical properties 
 are fully determined by the state $\rho$ of the measured system via the Born 
 rule.  If
the state of the system depends on some parameter $\lambda$, the
distribution $p(a|\lambda) = \hbox{Tr}[\rho_\lambda\, \ketbra{a}{a}]$ of the
outcomes of $A$ does depend on $\lambda$ as well. The CR
inequality (\ref{eq:cramerrao}) sets the lower bound to precision on any
estimation strategy for $\lambda$ based on the measurement of $A$.
Quantum estimation theory \cite{books,lqe1,lqe2,lqe3,lqe4} provides tools
to maximise the Fisher information over observables and to find the best
measurement to estimate a parameter.  The optimal measurement is defined
by the spectral decomposition of the so-called Symmetric Logarithmic 
Derivative (SLD) $L_{\lambda}$, which is implicitly defined through the equation:
\begin{equation}
    \frac12 (L_\lambda \rho_\lambda + \rho_\lambda L_\lambda) 
    \stackrel{\rm def}{=}\partial_\lambda \rho_\lambda\,,
    \label{eqsymm}
\end{equation}
where $\rho_\lambda$ is the quantum state, parametrized by an unknown
parameter $\lambda$, on which the measurement is performed. The quantum
Fisher information  is defined in terms of $L_\lambda$ as
\begin{equation}
    H(\lambda) \stackrel{\rm def}{=} \hbox{Tr}\llrr{\rho_\lambda L_\lambda^2}\,,
    \label{eq:QFI}
\end{equation}
and the ultimate bound to precision is set by the quantum CR 
inequality 
\begin{equation}
\hbox{Var} \lambda \geq \frac{1}{M H(\lambda)}\,.
    \label{eq:qcramerrao}
\end{equation}
In our case, the initial state $\rho_0 = \ketbra{\psi_0}{\psi_0} =
\ketbra{x_0}{x_0}$, and the evolved one $\rho_{\lambda} =
\ketbra{\psi_{\lambda}}{\psi_{\lambda}}$, where $\ket{\psi_{\lambda}} =
U_{\lambda} \ket{\psi_0} = \exp\llrr{-i \lambda G}\ket{\psi_0}$ are
pure. The expression for $\ket{\psi_\lambda}$ can be easily derived from
the spectral decomposition  \eref{eq:freeVal} and \eref{eq:freeSpec}.
 The generator $G$ is the self-adjoint operator ${\cal H}_0/\nu$, 
with ${\cal H}_0$ defined in
\eref{eq:H0}, i.e.,
\begin{equation}
G = -\frac{1}{2}\sum_{j=1}^{s-1}\ketbra{j+1}{j}+\ketbra{j}{j+1}.
\label{eq:G}
\end{equation}
The SLD takes the explicit form
\begin{equation}
    L_\lambda = \ketbra{\psi_\lambda}{\partial_\lambda \psi_\lambda}+ 
    \ketbra{\partial_\lambda \psi_\lambda}{\psi_\lambda}\,.
    \label{eq:SLDpure}
\end{equation}
Like for any unitary family of states, i.e. states that can be expressed as
$\ket{\psi_{\lambda}}=U_{\lambda}\ket{\psi_0}$, with $U_{\lambda}$ being an unitary transformation,
 the quantum Fisher information turns out to be
independent of the value of $\lambda=\nu t$, i.e. independent of the bare
coupling $\nu$ and on the interaction time. We have 
\begin{equation}
H = 4 \bra{\psi_0}G^2 \ket{\psi_0} -  
\llrr{\bra{\psi_0}G \ket{\psi_0}}^2,
 \label{eq:QFIpure}
\end{equation}
i.e. the quantum Fisher information is  proportional to the fluctuations of the 
generator on the initial
pure state \ket{\psi_0}.  
\par
The determination of the optimal measurement
through the spectral decomposition of $L_\lambda$ is straightforward. We
must, however, consider two families of initial conditions.  If the
excitation is initially located at one of the extremal sites of the
chain ($\ket{\psi_0} = \ket{1}$ or $\ket{\psi_0} = \ket{s}$), 
we have $H=1$, whereas for $\ket{\psi_0} = \ket{j},\
j\neq 1,s$ the quantum Fisher information is given by $H=2$. This result may be 
intuitively explained as
follows: the fluctuations of the generator $G$ acting on excitations
next to the boundaries of the chain are smaller than the fluctuations of
$G$ when it acts on an excitation that is free to move in both
directions ($\ket{j+1}$ or $\ket{j-1}$).  In order to determine the
optimal observable, we need the eigenvectors of $L_0$, with $L_\lambda
= U_\lambda L_0 U^\dag_\lambda$. It turns out that
$L_0$ for the initial excitation not being at the extremes of the chain (NE) 
admits the spectral decomposition:
\begin{align}
 &e_1^{\rm NE} = -\frac{1}{\sqrt{2}}, 
 \qquad e_2^{\rm NE} = \frac{1}{\sqrt{2}} \qquad e_j^{\rm NE} = 0,\mbox{ for }  j>2 \\
  &\ket{e_1^{\rm NE}} = \llrr{G - \frac{1}{\sqrt{2}}\mathbb{I}} \ket{\psi_0} \\
  &\ket{e_2^{\rm NE}} = \llrr{G + \frac{1}{\sqrt{2}}\mathbb{I}} \ket{\psi_0} \\
    &\ketbra{e_{\rm kern}^{\rm NE}}{e_{\rm kern}^{\rm NE}} = \mathbb{I} - \ketbra{e_1^{\rm NE}}{e_1^{\rm NE}} - \ketbra{e_2^{\rm NE}}{e_2^{\rm NE}}
\end{align}
whereas for the excitation at the extremes (E) we have
\begin{align}
    &e_1^{\rm E} = -\frac{1}{{2}}, \qquad e_2^{\rm E} = \frac{1}{{2}} \qquad e_j^{\rm E} = 0,\mbox{ for }  j>2 \\
    &\ket{e_1^{\rm E}} = \llrr{\sqrt{2} G - \frac{1}{\sqrt{2}}\mathbb{I}} \ket{\psi_0} \\
    &\ket{e_2^{\rm E}} = \llrr{\sqrt{2} G + \frac{1}{\sqrt{2}}\mathbb{I}} \ket{\psi_0}\\ 
    &\ketbra{e_{\rm kern}^{\rm E}}{e_{\rm kern}^{\rm E}} = \mathbb{I} - \ketbra{e_1^{\rm E}}{e_1^{\rm E}} - \ketbra{e_2^{\rm E}}{e_2^{\rm E}}.
\end{align}
As mentioned above, the eigenvectors of $L_\lambda$ are then 
given by:
\begin{equation}
    \ket{\psi_i^{\rm K}(\lambda)} =
    U_\lambda \ket{\psi_i^{\rm K}},
    \label{eq:evolEigenL}
\end{equation}
with ${\rm K}={\rm NE},{\rm E}$.
The spectral decomposition of $L_\lambda$ 
defines an admissible observable for any
value of the parameter which, however, may be hard
to implement in a realistic scenario.  A question thus arises about the
performances of other kind of measurements, which may correspond to
feasible interaction schemes, at least in principle. In the following,
we analyze estimation procedures based on local measurements and 
on Feynman probes and assess their performances in terms of
precision, that is, we compare their Fisher information to the quantum
Fisher information.
\section{Local measurement}\label{locmesS}
The effective coupling parameter $\lambda$ is the transition rate 
for the excitation to  move to an adjacent site.
For example, it tells us the rate at which the particle leaves
 its initial position $x_0$. This suggests a
simple measurement to infer the value of $\lambda$: we place the
excitation initially at a given site $x_0$ and, after the chosen 
interaction time, 
we test via local measurement at $x_0$  whether the excitation 
is still there or not.  The
information about the unknown parameter $\lambda$ obtained through 
this kind of measurement may be quantified by the classical Fisher
information of the associated distribution, which consists of a Bernoulli
trial with success probability $P^L_\lambda(x_0) = 
\left|\langle\psi_\lambda|x_0\rangle\right|^2$. The 
Fisher information for the local measurement is thus given by
\begin{align}
F_{x_0}^L(\lambda;x_0)= \frac{\left[\partial_\lambda 
P^L_\lambda(x_0)\right]^2}{
P^L_\lambda(x_0)
[1-P^L_\lambda((x_0)]}.
    \label{eq:fisherInfo1}
\end{align}
The test measurement may be, of course, performed at a different 
site $m \neq x_0$ (still placing the excitation initially at $x_0$).
The corresponding Fisher information then reads
\begin{equation}
F_m^L(\lambda;x_0)= \frac{\left[\partial_\lambda P^L_\lambda(m|x_0)
\right]^2}
{P^L_\lambda(m|x_0) [1-P^L_\lambda(m|x_0)] }   
    \label{eq:fisherInfo2}
\end{equation}
where
\begin{equation}
P^L_\lambda(m|x_0)
= \left |\langle\psi_\lambda|m\rangle\right |^2.\label{prLM}
\end{equation} 
The analytic expression of the probabilities 
$P^{L}_\lambda(m|x_{0})$ is cumbersome but can be straightforwardly derived from \eref{eq:freeVal} and \eref{eq:freeSpec}, so we do not report it here.
\par
\begin{figure}[tb]
\centering
\subfigure[ \,\,$x_0=1$]
{\includegraphics[width=0.90\columnwidth]{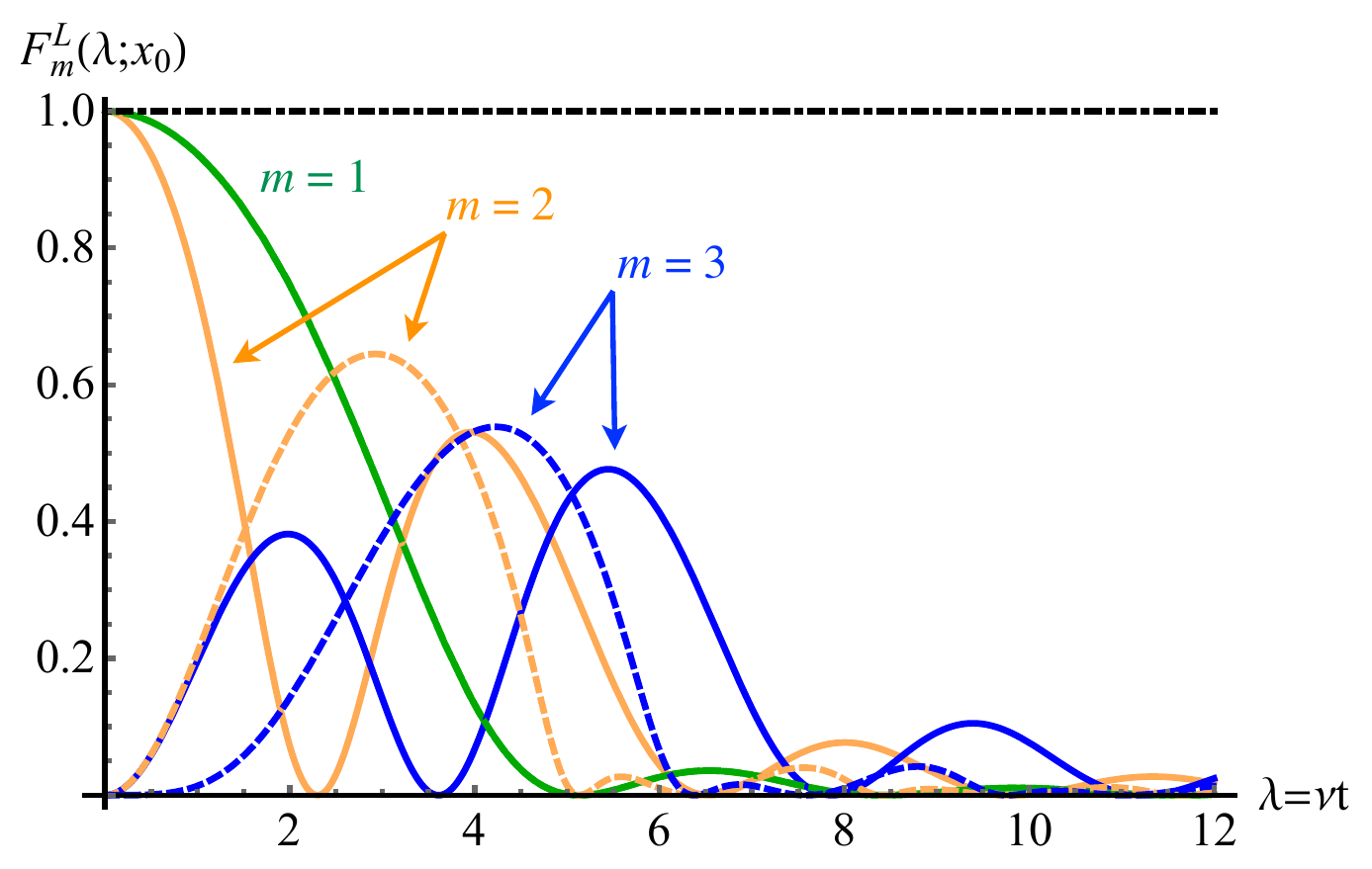}} \label{fig:fexa}
\subfigure[ \,\,$x_0=2$]{
\includegraphics[width=0.90\columnwidth]{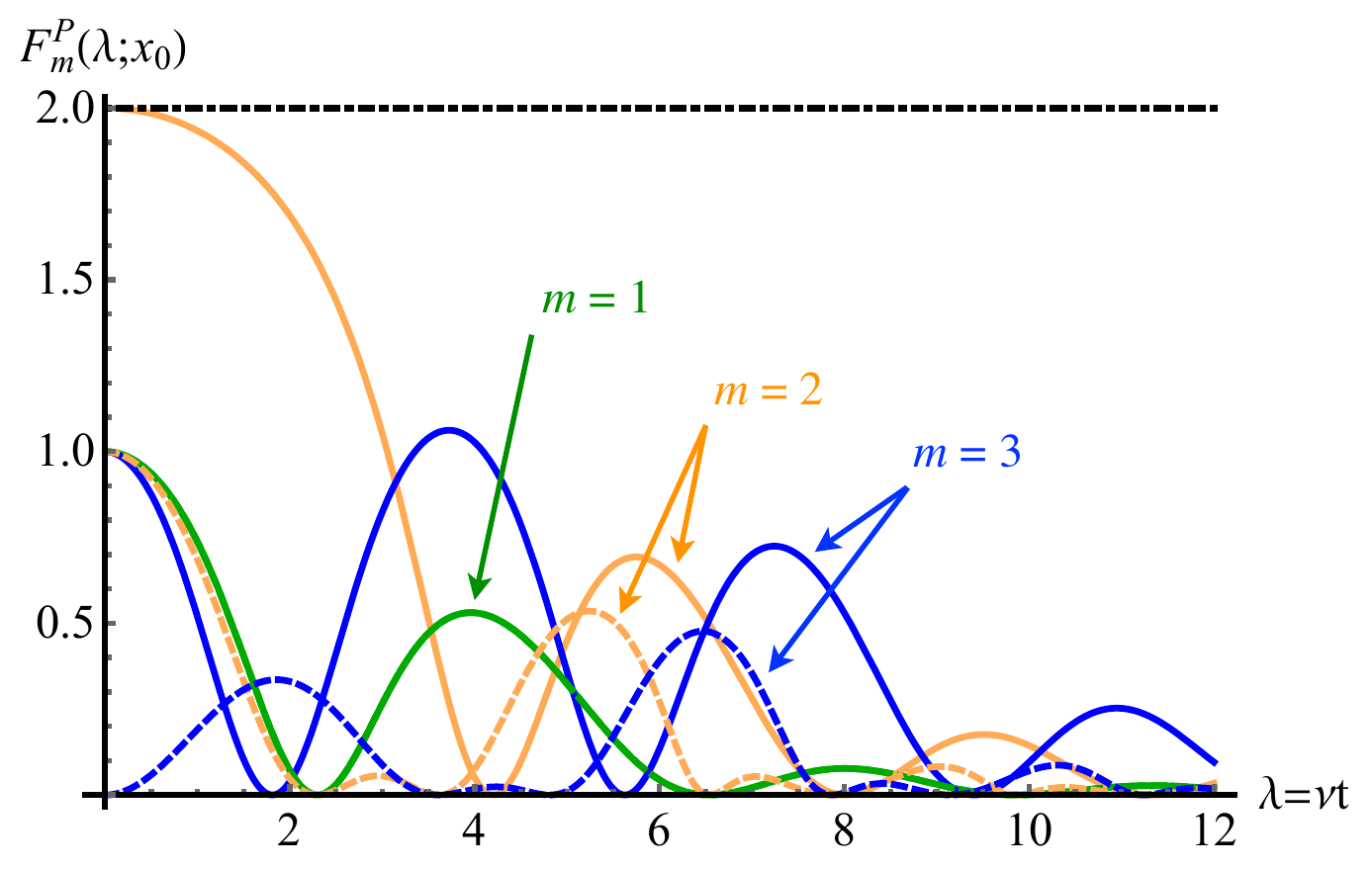}} \label{fig:fexb}
\caption{(Color online)  Fisher information for the local measurement
$F_m^L(\lambda; x_0)$ (solid lines) and for the Feynman probe
$F_m^P(\lambda; x_0)$ (dashed lines) as a function of $\lambda = \nu t$. The data refer to a chain of $s=10$ spins for different values of the measured or plugging site: $m=1$ (green  lines ),  $m=2$ (orange  lines), and   $m=3$ (blue lines). (a) Initial condition
set to $x_0=1$;  (b) $x_0=2$. In both frames the dot-dashed black line represents the quantum Fisher information.}   \label{fig:gexab}
\end{figure}

In Fig.\ref{fig:gexab} we show the evolution of $F_m^L(\lambda;x_0)$
as a function of $\lambda$  for three different values of the measured site
 $m = 1,2,3$ and two different initial
conditions: $x_0=1$ [Fig. 1(a)] and $x_0=2$  [Fig. 1(b)]. At $t=0$ the
Fisher information $F_{x_0}^L(\lambda;x_0)$ of the observable
$\ketbra{x_0}{x_0}$ saturates the quantum Fisher information; this is a general fact: given an
arbitrary initial condition $\ket{\psi_0} = \delta_{x,x_0} \ket{x}$, the
most efficient projective measurement is the projector
\ketbra{x_0}{x_0}. The Fisher information $F_m^L(\lambda;x_0)$, on the
other hand, does not saturate the quantum Fisher information; its maximum, achieved after 
an interaction time
proportional to $|m-x_0|/\nu$, is, in general, well below the quantum Fisher information
threshold. 
\section{Feynman Probes}\label{FePS}
\begin{figure}[h]
\centering
\includegraphics[width = 0.70 \columnwidth]{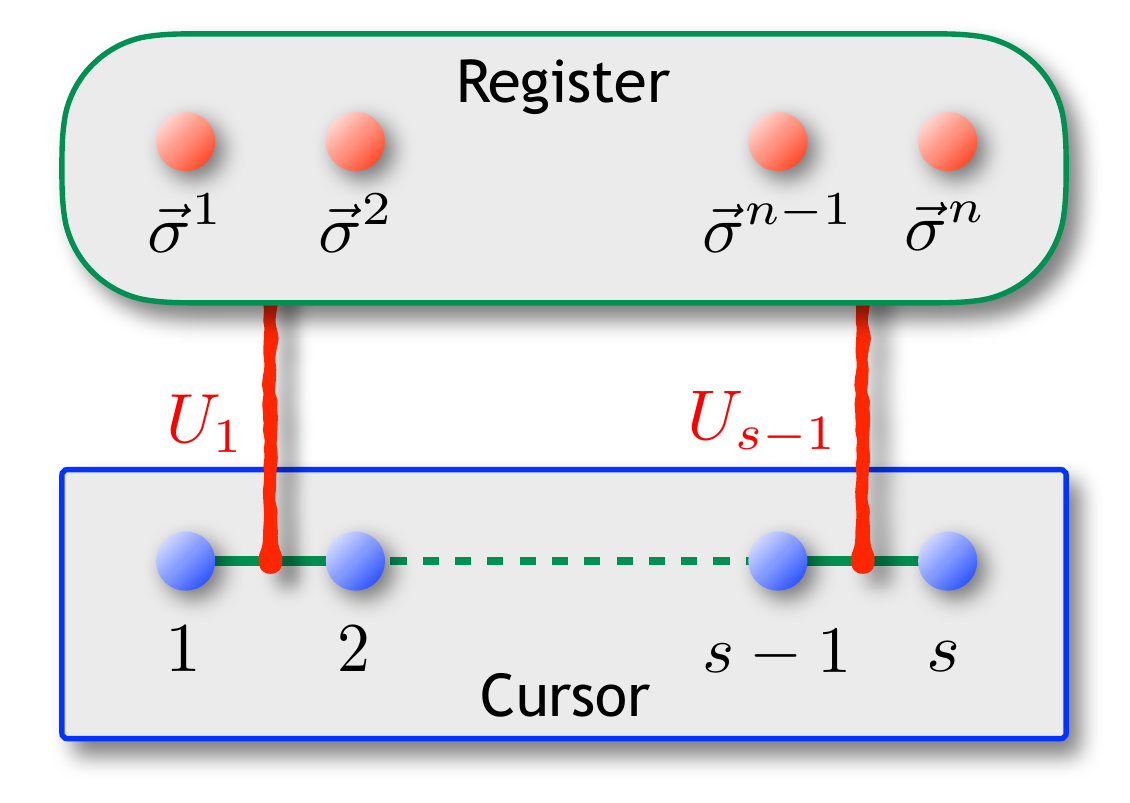}
\vspace{-0.3cm}
\caption{A schematic representation of the Feynman quantum computer}
\label{fig:feynMachine}
\end{figure}

Feynman's quantum computer \cite{feyn87} consists of two logically 
separated parts; one part, the \emph{clock}, is an excitation moving 
along a lattice. The second part, the input-output register, 
is a collection of additional degrees of freedom, say $n$ spin-1/2 
particles $\vec{\sigma}^j$, $j=1,2,
\ldots,n$. The overall system is governed by the time-independent
Hamiltonian:
\begin{eqnarray} \label{eq:feyn}
 {\cal H}_F = -\frac{\nu}{2}\sum_{j=1}^{s-1} \ketbra{j+1}{j} \otimes U_j 
 + \ketbra{j}{j+1}\otimes U_j^{-1}.
\end{eqnarray}
Each term of the Hamiltonian involves two nearest neighbor sites of 
the clock and a self-adjoint or unitary operator $U_j$ acting on 
the register.  The ordered product $U_{s-1} \ldots U_2 U_1$ realizes 
some input-output transformation that the computing device is expected 
to accomplish. \Fref{fig:feynMachine} shows the architecture of 
the machine.
Because of the properties of the Hamiltonian ${\cal H}_F$, the position 
of the excitation in the clock, i.e., along the chain, uniquely 
determines the state of the register. This fact has interesting 
consequences. Let us consider the overall machine initially prepared in 
the state $\ket{\psi_0}=\ket{1}\otimes\ket{R_1}\equiv \ket{1,R(1)}$, i.e., 
with the excitation located at the beginning of the chain and 
the register in the pure initial state $\ket{R_1}$. Then 
the set
\[
\mathcal{B}(\psi_0)=\{ \ket{1,R_1}, \ket{2,R_2},\ldots,\ket{s,R_s}\},
\]
where $\ket{R_j} = U_{j-1} \ldots U_2 U_1 \ket{R_1}$, 
constitutes an orthonormal \emph{computational basis}, often referred
to as the \emph{Peres basis} \cite{peres85}, for the region of  
the Hilbert space  visited by the evolved state $\ket{\psi_t} = 
\exp(-i {\cal H}_F t)\ket{\psi_0}$. This basis may be defined constructively 
for any choice of the initial condition \ket{\psi_0}. We refer to the 
space spanned by the Peres basis as the \emph{computational 
subspace}. In particular, if upon measurement the clock is found 
at the rightmost site of the chain, the register collapses to the 
output state $\ket{R(s)} = U_{s-1} \ldots U_2 U_1 \ket{R(1)}$. 
Before discussing the kinematics of the clock, we point out that 
the sole effect of the interaction of the clock with the register 
of $n$ spins is the appearance of a degeneracy of order $2^n$ in 
the spectrum $\{e_k(\nu)\}_{k=1}^s$ of the tight-binding (clock) 
Hamiltonian 
\begin{equation*}
    {\cal H}_0 
    = -\frac{\nu}{2} \sum_{j=1}^{s-1} \ketbra{j+1}{j} + \ketbra{j}{j+1},
\end{equation*}
i.e. the Hamiltonian \eref{eq:H0}. Once an initial condition of the 
form $\ket{\psi_0}=\ket{1,R(1)}$ has been set, however, the spectrum of 
the Hamiltonian ${\cal H}_F$, restricted to the computational subspace, is no 
longer degenerate. In this subspace, the eigenvector
\begin{align}
\ket{v_k} &= \sqrt{\frac{2}{s+1}}\sum_{j=1}^s 
\sin\left (\frac{k \pi j}{s+1} \right) \ket{j,R(j)}.
\end{align}
corresponds to each eigenvalue $e_k(\nu)$.
The properties of the Feynman quantum computer have been extensively
discussed \cite{def06, def13}. The most relevant property of the Feynman
machine that we want to exploit here is the entanglement between the
clock and the register. The idea is to gain information about some
physical parameter characterizing the clock by performing suitable
measurements on the register alone. To this aim, we consider a
streamlined version of the Feynman machine.  Indeed, we consider a register made
up of a {\em single} two-level system,  which we refer to as the \emph{probe}. 
The probe is initialized, without loss of generality, into the 
eigenstate of $\sigma_z$ belonging to the 
eigenvalue $+1$, or the \emph{up} state \ket{\uparrow}. All the 
operators $U_j$ but the $m$-th one, $ 1 \leq 
 m \leq s-1$, are set
to $\mathbb{I}$, whereas $U_m = \sigma_x$. 
In this setting, the Feynman Hamiltonian \eref{eq:feyn} reads 
\begin{align}
    {\cal H}_F(m) &=
    -\frac{\nu}{2}\sum_{\substack{j=1 \\ j\neq m}}^{s-1} \ketbra{j+1}{j} + \ketbra{j}{j+1} 
    + \nonumber \\
    &-\frac{\nu}{2} \llrr{\ketbra{m+1}{m}\otimes \sigma_x + 
    \ketbra{m}{m+1} \otimes \sigma_x}. 
    \label{eq:intTerm}
\end{align}
If the clock is initially at a site $x_0\leq m$, 
the Peres basis for the system is 
\begin{equation}
\ket{1,\uparrow},\ldots,\ket{m,\uparrow},
\ket{m+1,\downarrow},\ldots,\ket{s,\downarrow}.
    \label{eq:basis}
\end{equation}
Upon a projective measurement $\mathbb{I}\otimes
\ketbra{\uparrow}{\uparrow}$ of the $\sigma_z$ component of the probe,
the evolved state $\ket{\psi_\lambda} $ collapses
into a state with support in either the
$\Span\llrr{{\ket{1,\uparrow},\ldots,\ket{m,\uparrow}}}$ or
$\Span\llrr{\ket{m+1,\downarrow},\ldots,\ket{s,\downarrow}}$ subspace of
the Hilbert space of states.
\par
As a matter of fact, the Feynman structure provides a nonlocal alternative to
characterize the qubit chain. Instead of measuring whether the
excitation has left its initial position $x_0$, or reached a target one 
$x_m$, we can measure the $\sigma_z$ observable of the probe qubit, 
 which we will now refer to as  the 
\emph{Feynman probe}.
The Fisher information associated with such measurement is given by
\begin{align}
 F_m^P(\lambda;x_0)=\frac{\left[\partial_\lambda 
 P_\lambda(\uparrow|m,x_0) \right]^2}{P_\lambda(\uparrow|m,x_0)
 [1-P_\lambda(\uparrow|m,x_0)]}
\end{align}
where 
\begin{equation}
P_\lambda(\uparrow|m,x_0)=|\langle \psi_{\lambda}|\uparrow\rangle|^2\label{prFP}
\end{equation}
is the probability of measuring the Feynman probe in the state
$\ket{\uparrow}$, when it is plugged into the $m$-th site of the chain
and the excitation is initially located at site $x_0$. Because of the
probe-system entanglement, we have  
\begin{equation}
P_\lambda(\uparrow|m,x_0)=
\sum_{x=1}^m |\langle \psi_{\lambda}|x\rangle|^2.
\label{eq:fpp}
\end{equation}
The behavior of $ F_m^P(\lambda;x_0)$ for three different 
plugging sites $m=1,2,3$ 
and two different initial positions $x_0=1,2$ is
shown in Fig.\ref{fig:gexab}. For $x_0=1$ and $m=1$ a Feynman probe 
provides the same information as a local measurement, i.e.
$F_1^L(\lambda;1) =F_1^P(\lambda;1)$. For $m=2$, $F_2^L(\lambda;1)$
still saturates the quantum Fisher information at $t=0$, whereas $F_2^P(\lambda;1)$ does not. 
For $m \geq 3$, the maximum Fisher information 
for the Feynman probe is typically larger than the
Fisher information of the corresponding local measurement, whereas 
for $x_0 >1$, local measurements typically carry more information than the
Feynman probe for any value of $m>1$.
\par
Overall, from the point of view of the efficiency of the measurement
scheme alone, our results show that the ultimate bounds to the precision
of Feynman probes are enhanced compared to those of local measurements
when the excitation is initially located at the boundary of the chain
and $m >2$.  On the other hand, a proper comparison should be made in
terms of an actual estimation strategy and thas is the scope of the
next section.
\par
Notice that the implementation of the Feynman Hamiltonian of Eq.
(\ref{eq:intTerm}) involves a three-spin interaction. This may be
challenging from the experimental point of view. On the other hand,
promising proposals based on cold atoms in optical lattices have already
been discussed \cite{duan03,plenio04,cosco16}. 
%
\begin{figure}
 \centering
 \includegraphics[width=0.90\columnwidth]{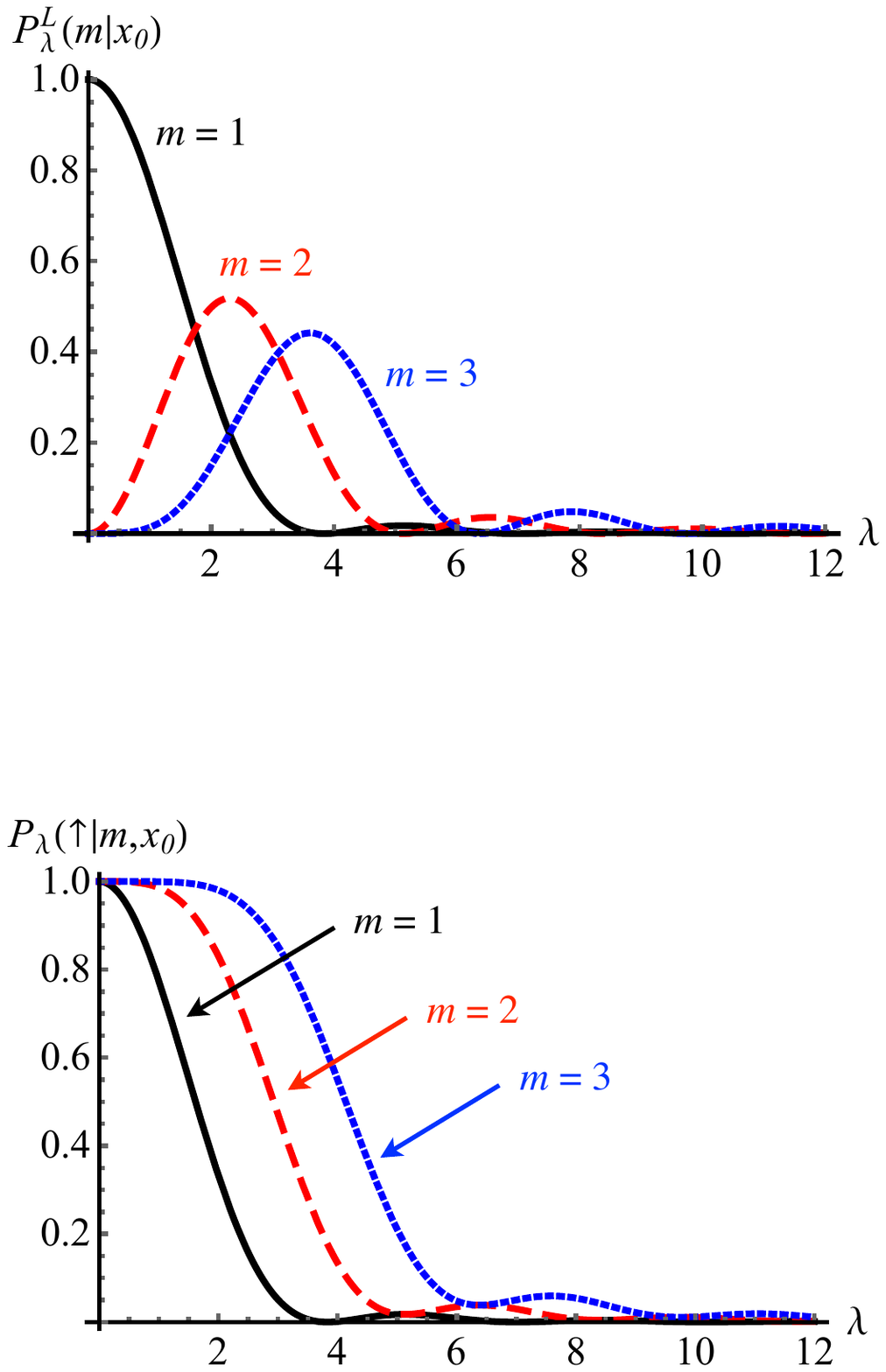}
 \includegraphics[width=0.90\columnwidth]{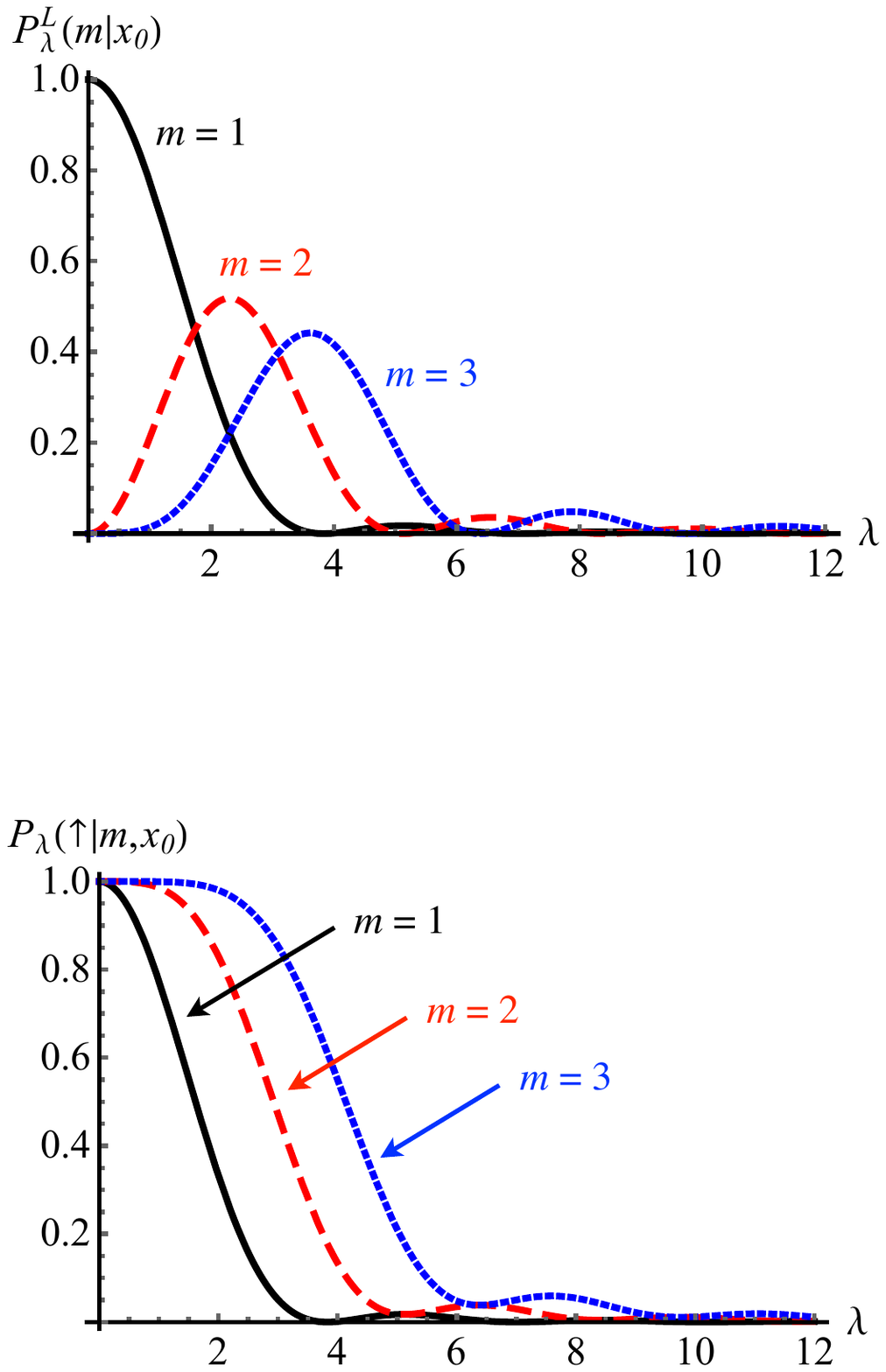}
 \vspace{-0.5cm}
\caption{(Color online) $P_{\lambda}^L(m|x_0)$ (top plot) and $P_{\lambda}(\uparrow|m, x_0)$ (bottom
plot) as  defined in Eqs. \eqref{prLM} and \eqref{prFP} respectively, as a function of the parameter $\lambda$. The excitation is
initially localized in the first site  $x_0=1$ of a chain of length
$s=10$. Three different values of $m$ are considered: $m=1$, (solid black line),
$m=2$ (dashed red line) and $m=3$ (blue dotted line).} \label{prob_dist}
\end{figure}
\section{Bayesian estimation}\label{bayesS}
\begin{figure*}[tb]
\centering
\includegraphics[width=0.33\textwidth]{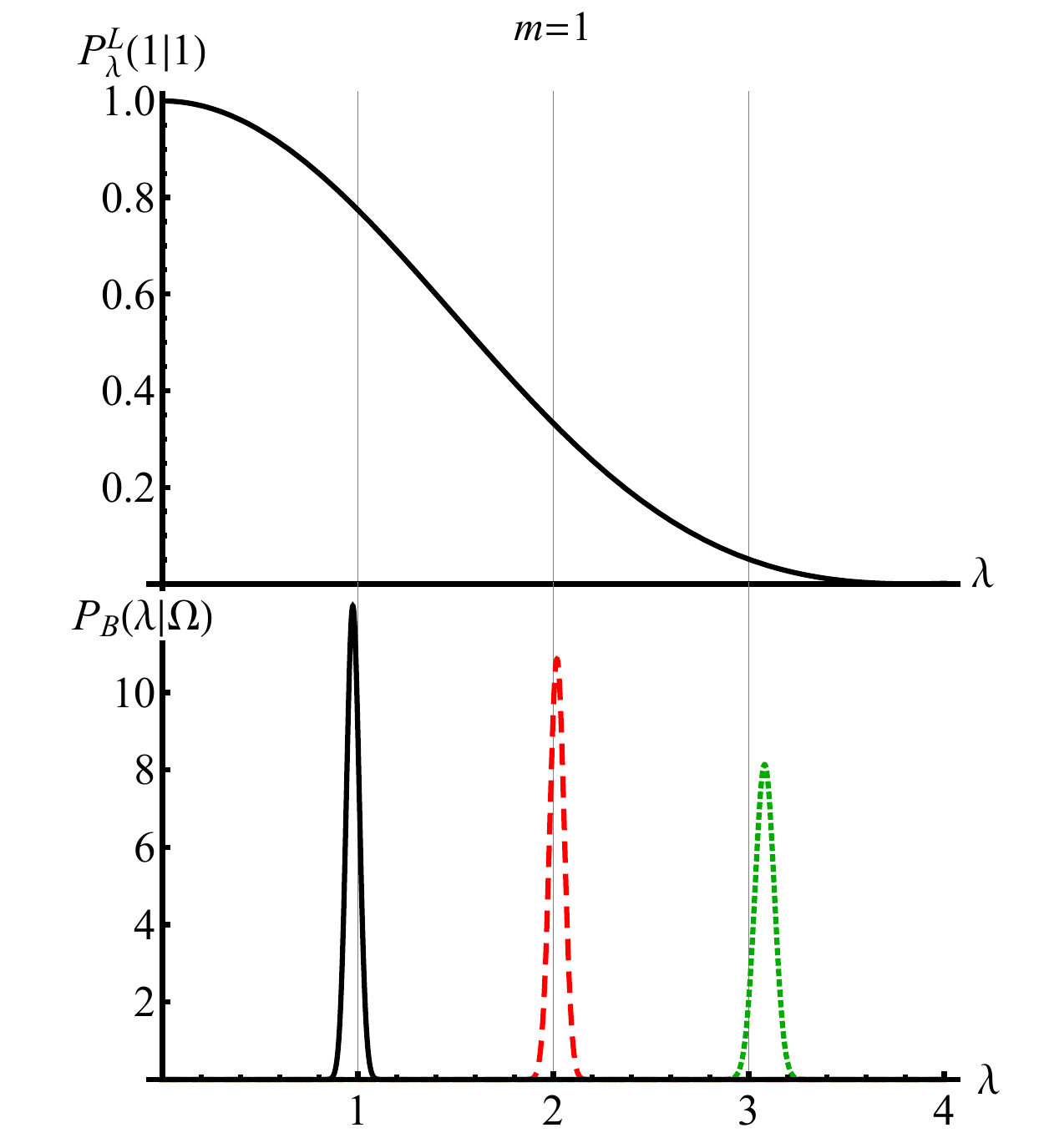}\hspace{-3.mm}
\includegraphics[width=0.33\textwidth]{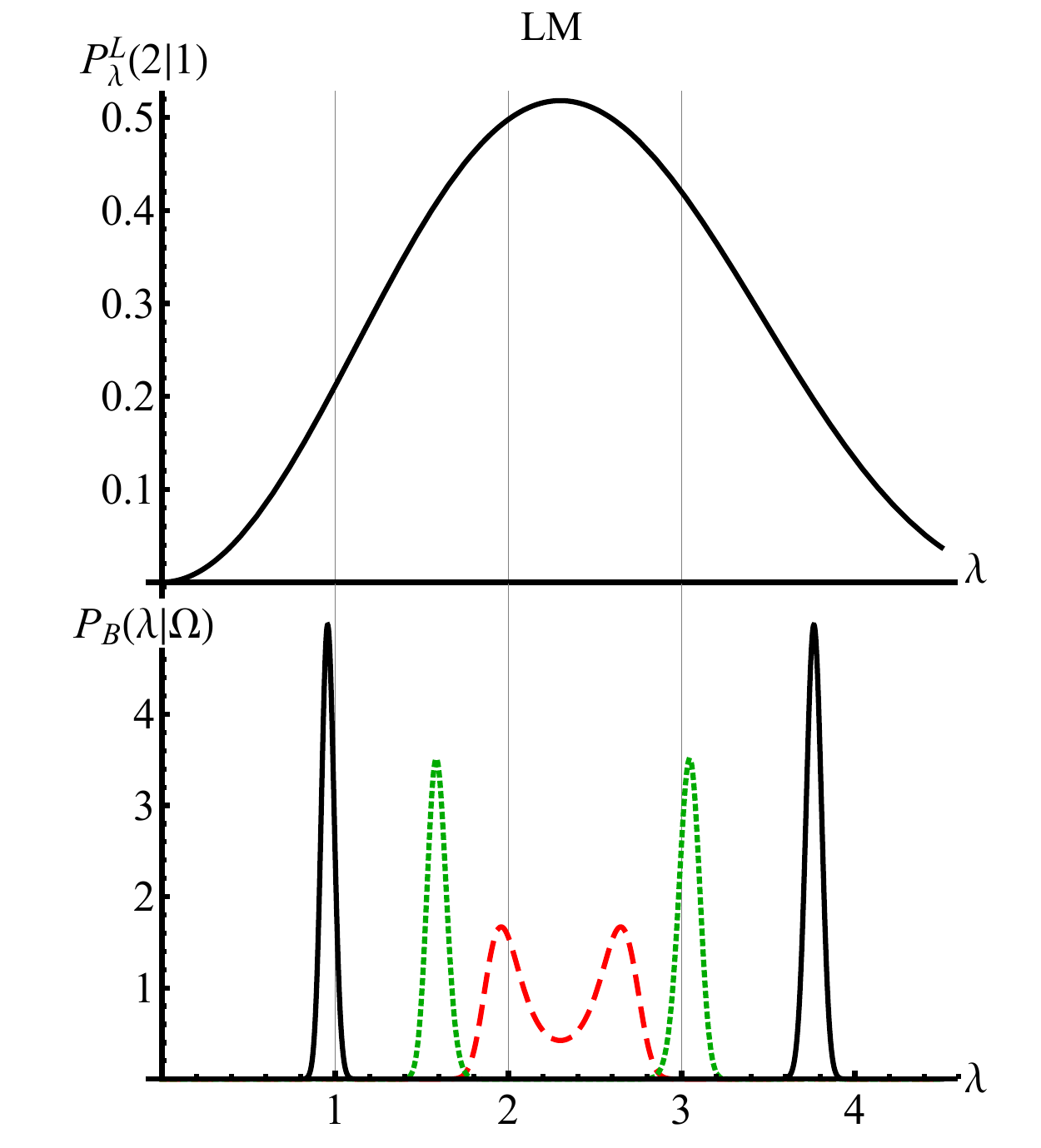}\hspace{-3.mm}
\includegraphics[width=0.33\textwidth]{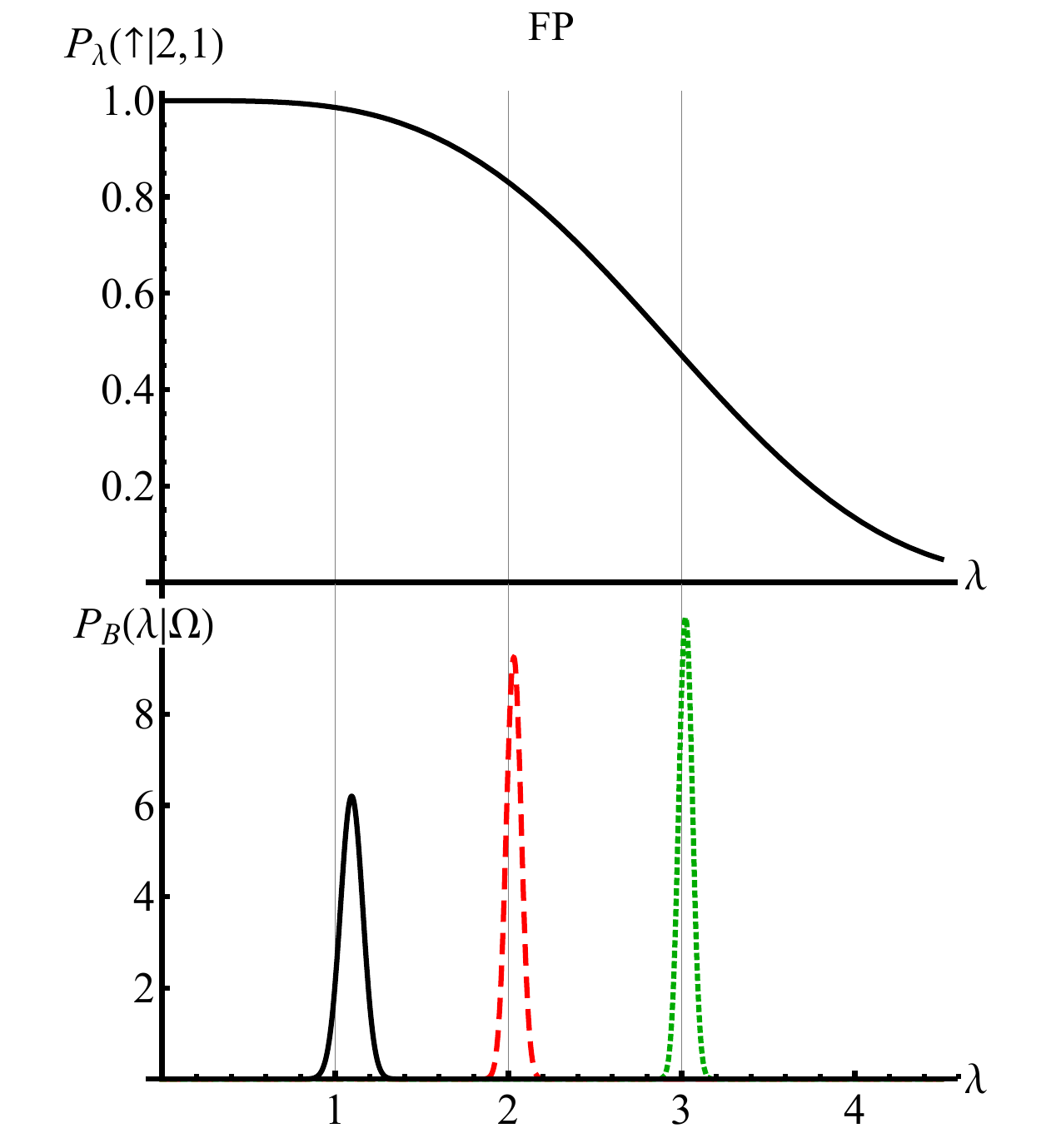}
\vspace{-0.3cm}
\caption{(Color online) (Top) Success probability in a single trial for
LM and FP as defined in Eqs. \eqref{prLM} and \eqref{prFP},
respectively; (bottom) Bayesian probability $P_B(\lambda|\Omega)$
defined in Eq. \eqref{BayProb}. The plots refer to simulations of (left
column) LM on the site $m=1$; (middle column) LM on the site $m=2$ and
(right column) FP coupled to site $m=2$. In the bottom plots, we use the
true values $\lambda_T=1$ (solid black line), $2$ (dashed red line) and $3$
(dotted green line).  The vertical lines are a guide for the eye marking the
values of $\lambda_T$.}  \label{be_x1}
\end{figure*}
%
Classical and quantum CR theorems pose bounds to the precision
of {\em any} unbiased estimator of the parameter of interest. However,
no recipes are given to find optimal estimators saturating the classical
bound. Therefore, in order to properly compare the performances of the
local measurement (LM) and of the Feynman probe (FP), and to assess them against
the ultimate quantum bound, we employ a Bayesian estimation strategy for
the parameter $\lambda$, starting from a numerically simulated set of
experimental data. Indeed, Bayesian estimators are known to be
asymptotically optimal; that is, they saturate the CR bound for
large data samples. Hereafter, we fix the initial position of the
excitation in the first site $\ket{x_0}=\ket{1}$.  The key ingredient in
the Bayesian estimation is the Bayes theorem, which we can
rewrite as:
\begin{equation}
 P_B(\lambda|\Omega)=\frac{P(\Omega|\lambda) P(\lambda)}{\int P(\Omega|\lambda') P(\lambda')d\lambda'}
 \label{BayProb}
\end{equation}
where $P_B(\lambda|\Omega)$ is the {\it a posteriori} Bayesian
probability distribution  of the parameter  $\lambda$ given 
 the set of experimental data $\Omega$, whereas
$P(\lambda)$ is the {\it a priori} probability distribution of $\lambda$. We assume we do not have any prior knowledge about
the estimable parameter, so we can consider a flat distribution for
$P(\lambda)$. The quantity $P(\Omega|\lambda)$ is the likelihood of
obtaining the set of experimental data $\Omega$ when the true value of
the parameter is $\lambda$. In our case, we have $M$ identical repeated
Bernoulli trials, each of which  is characterized by a success probability
$p_\lambda$, then the likelihood of having $N_0$ successes out of $M$ is
given by
\begin{equation}
 P(\Omega|\lambda)=p_\lambda^{N_0}\,\left(1-p_\lambda\right)^{M-N_0}.
\end{equation}
If the performed measures are local measurements on a
site $m$, then $p_\lambda=P^L_\lambda (m|x_0)$, 
while in the case we measure  the Feynman probe coupled 
to sites $m$ and $m+1$, we have $p_\lambda=P_\lambda(\uparrow|m,x_0)$.
Once we reconstruct the Bayesian probability distribution \eqref{BayProb}, we can estimate the parameter as the expectation of 
the random variable $\lambda$:
\begin{equation}
 \hat{\lambda}=\int\, \lambda P_B(\lambda|\Omega) d\lambda.
 \label{best}
\end{equation}
Accordingly, the variance of such an estimator is computed as:
\begin{equation}
 \sigma^2[\hat{\lambda}]=\int\, [\lambda-\hat{\lambda}]^2 P_B(\lambda|\Omega) d\lambda.
\end{equation}
The probability $P_{\lambda}^L(m|x_0)$ of measuring the excitation on site
$m$   is shown in Fig. \ref{prob_dist} for three different values of
$m$, together with the probability $P_{\lambda}(\uparrow|m,x_0)$ of measuring
the FP in the state $\ket{\uparrow}$ when it is plugged between sites
$m$ and $m+1$.  If we perform a local measurement on the first site
$m=1$, then $P_{\lambda}^L(1|1)=P_{\lambda}(\uparrow |1,1)$ and the two
strategies are equivalent. The behavior of this probability is
shown in Fig. \ref{be_x1}. The success probability for each of the $M$
trials has a monotonic behavior, giving rise to a single-peak Bayesian
probability distribution. The width of the peak is related to the
variance of the estimator. 
If we measure any site other than the first one, then the local
measurement and the FP strategy lead to dramatically different results,
as shown in Fig. \ref{be_x1}.  We first notice that the 
probability $P_{\lambda}^L(m|1)$ is not invertible as a function of
$\lambda$ except for the value corresponding to its maximum. On the
other hand, it is possible to find a value $\lambda_{max}$ such that  $P_{\lambda}(\uparrow|m,1)$ is invertible in the region
$[0,\lambda_{\max}]$ of the parameter space.
This has profound consequences on the reliability of the two estimation
procedure, because it affects the shape of the {\it a posteriori}
probability distribution for  estimating the parameter. In fact, the
Bayesian probability built according to Eq. \eqref{BayProb} has
two peaks for a local measurement procedure, corresponding to the two
values of the parameter $\lambda$  which give the same probability
$P_{\lambda}^L(m|x_0)$,  and a single maximum for the FP estimator.
Indeed, the probability distribution
$P_{\lambda}^L(m|x_0)$ is non-monotonic, showing a maximum corresponding to
the only value of the parameter $\lambda$ that can be estimated without
any prior knowledge about $\lambda_T$. For any other values of
$\lambda_T$, the local measurement strategy fails to uniquely identify a
single solution for the estimation procedure and a two-peak probability
distribution $P_B(\lambda|\Omega)$ is obtained. For $\lambda_T$
approaching the maximum of the
distribution, the two peaks start to merge, leading to a broad 
probability distribution. On the contrary, the FP success probability
keeps its monotonic behavior in the whole parameter region and thus 
Bayesian inversion strategy always leads to a single
solution, within the variance of the estimator. 
Except for the case where the local measurement is performed on the
first node $\ket{1}$ of the lattice, a presence measurement on any other
site will give a {\itshape a-posteriori} probability distribution with
two peaks, i.e.  the Bayesian estimation procedure identifies two
possible solutions $\lambda_1$, $\lambda_2$. Since there is no way to
discriminate between $\lambda_1$ and $\lambda_2$ without any {\itshape
a-priori} knowledge, the local measurement cannot provide a 
reliable estimation of the coupling strength $\nu=\frac{\lambda}{t}$. Measuring
the state of the FP, on the other hand, allows us to correctly infer
the value of $\lambda$, within an error given by the width of the
single-peak reconstructed Bayesian probability distribution.  Once we
fix the time at which we perform the measurement $t_M$, the coupling
constant is easily identified as $\nu=\lambda/t_M$. 
\par
\begin{figure}[h!]
\centering
\includegraphics[width=0.9\columnwidth]{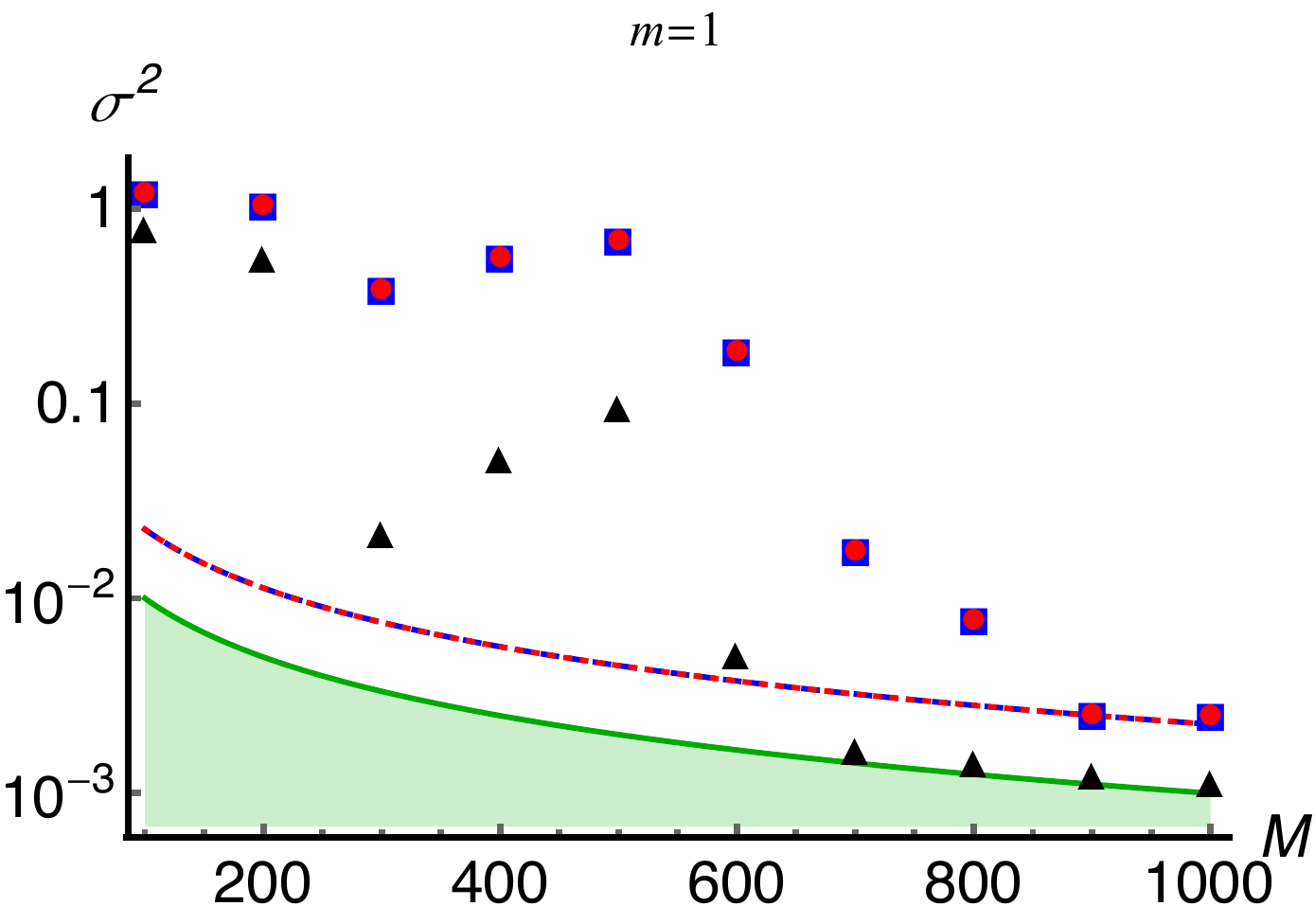}\vspace{5mm}
\includegraphics[width=0.9\columnwidth]{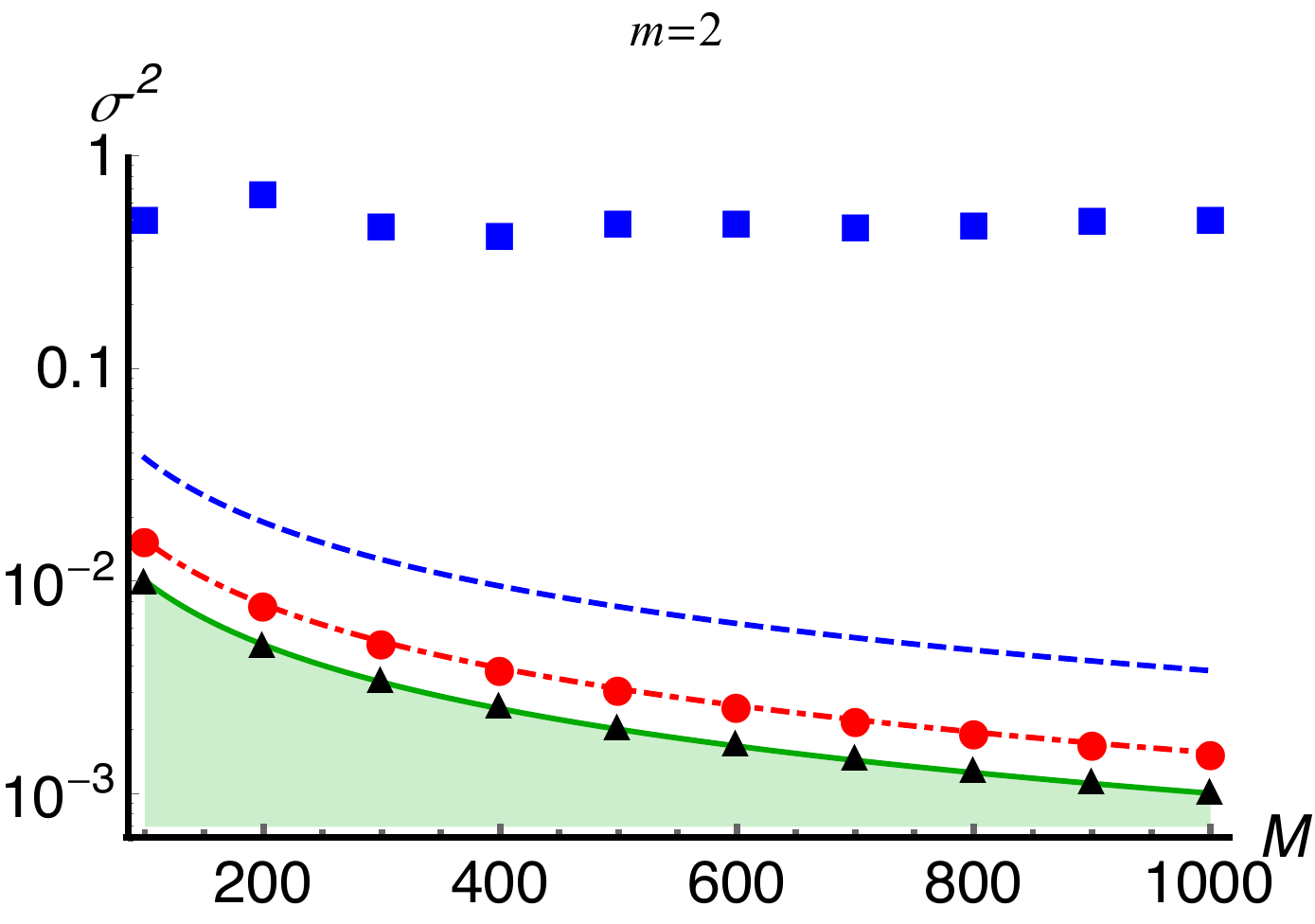}\vspace{5mm}
\includegraphics[width=0.9\columnwidth]{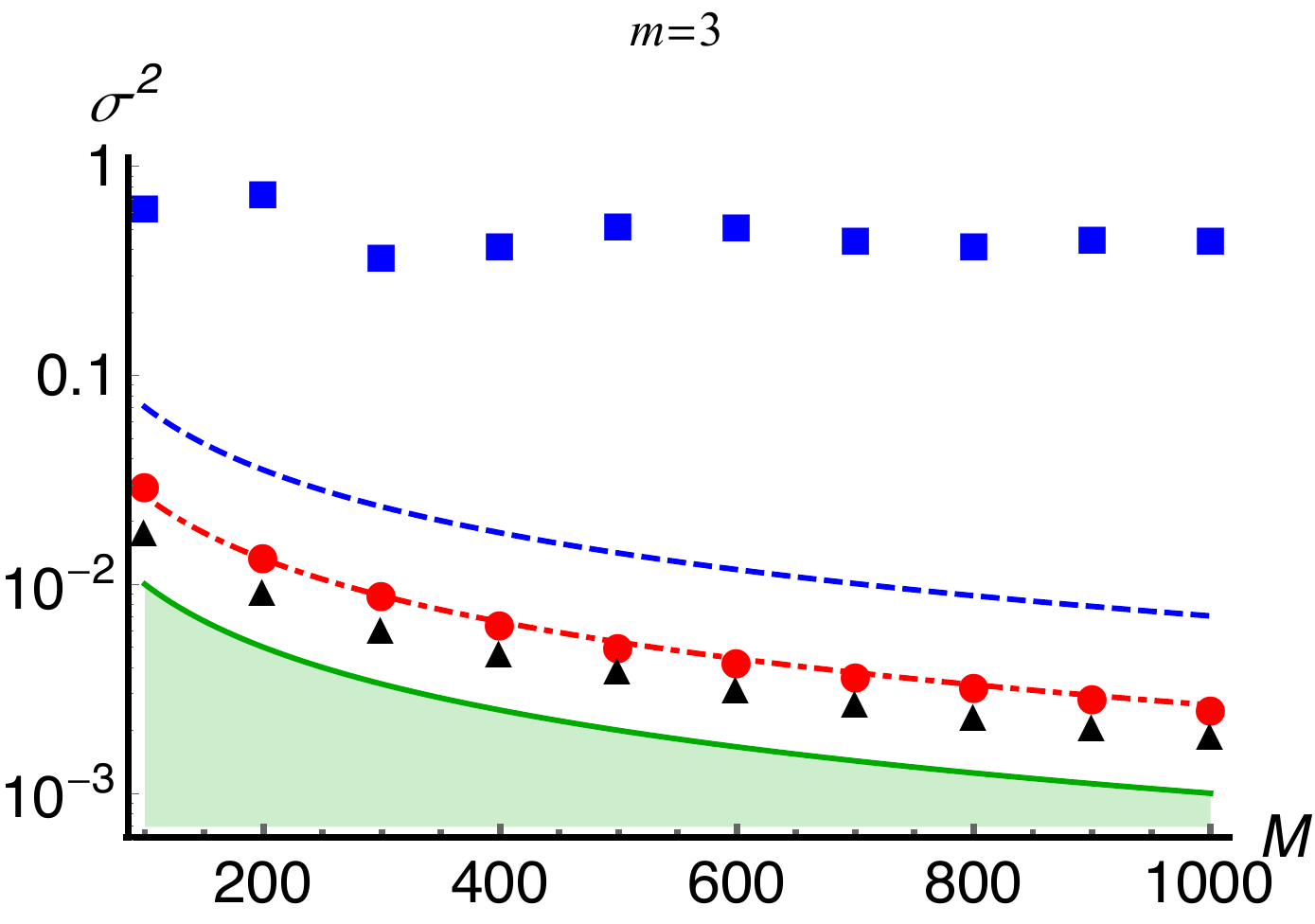}
\caption{(Color online) Variance for the LM (blue squares), FP (red circles) and combined
FP+LM (black triangles)  estimators as a function of the number $M$ of repeated simulated
measurements, for $\lambda_T=3$. Each plot refers to a different value of the measured (or plugging) site $m$. The  
blue dashed line and the red dot-dashed one highlight the limit 
imposed by the CR bound for LM and FP
measurements respectively. The green shaded area 
delimits the region forbidden by the quantum CR bound.} 
\label{crf}
\end{figure}

The Bayesian FP estimator saturates the CR
bound \eqref{eq:cramerrao}, as shown in  Fig. \ref{crf}. However, as we
have already mentioned, neither the LM nor the FP allows us to saturate
the quantum CR bound. In order to improve precision 
we may use a two-step scheme, in which
we first employ the Feynman probe and Bayesian estimation and then 
use the posterior
distribution for $\lambda$ as a prior for Bayesian estimation 
using local measurements. This
procedure allows us to build a more precise estimator, with a smaller
variance compared to those obtained using solely local measurements
or Feynman probes.  The variance $\sigma^2$ obtained from a set of $M$
simulated experiments which employ this LM+FP scheme is shown in Fig.
\eqref{crf} (black dots). As it is apparent from the plot 
the two-step scheme performs better than the 
other measurements schemes and it allows one to achieve, 
in some cases, the quantum CR bound, thus representing an optimal
procedure to characterize linear chain of qubits.
\par
\par
\section{Conclusion and outlook}\label{conlS}
In this paper we have introduced a novel estimation procedure based 
on Feynman probes for the characterization of linear qubit chains
made of strongly interacting two-level systems.  A Feynman probe is 
a single-qubit register coupled to the chain of qubits, 
which is initially prepared in  a single-excitation state. 
The probe is dynamically entangled to the excitation moving along the 
linear chain and by measuring the state of the probe 
it is possible to extract information on the value of 
the coupling constant.
\par
First, we evaluated the quantum Fisher information associated with
the coupling parameter, thus determining the ultimate precision of any
estimation procedure.  We then showed that local measurements, i.e.
 measuring the presence of  the excitation on a single site, provide
optimal characterization of $\lambda$ for small values of the coupling. 
In particular, optimal estimation is obtained when the 
local measurement is performed at site $x_0$, where the
excitation is initially localized.
In this regime the CR bound may be attained 
by Bayesian estimation, for a large number of repeated measurements.
On the contrary, for larger values of the coupling $\lambda$, 
i.e. for strongly coupled chains of qubits,
a Bayesian local strategy does not lead to a consistent estimate,
because the  {\it a-posteriori} probability distribution 
shows two peaks. In this regime, 
Feynman probes provide a consistent
Bayesian estimator that saturates the CR bound, i.e. it achieves
efficient characterization of the qubit system.  We concluded that
characterization by Feynman probes represents  a suitable estimation
strategy for a strongly coupled qubit chain. 
\par
Finally, we suggested a two-step measurement scheme where 
both FP and LM are employed one after the other to estimate the 
coupling $\lambda$. In the first step, FP is used to infer the 
coupling: the resulting distribution is then used as an a-priori 
distribution for a Bayesian LM estimation $\lambda$. The overall
precision may achieve the quantum CR bound.
Our results provide an
alternative route to characterize qubit systems and confirm the
relevance of nonlocal measurements, which have already been suggested as
a convenient toolbox for quantum circuits based on trapped ions
\cite{ti1,ti2} and  supeconducting qubits \cite{nig13,blu16}. Feynman probes could  also be employed to estimate the current in out of equilibrium quantum wires \cite{benenti09,clark13} or the amount of disorder in linear lattices \cite{lahini08,tama09}.
\par
As we pointed out, cold atoms in optical lattices are promising systems for the realization of the three-spin interaction needed by Feynman probes. Our results confirm the
interest of these systems and may foster future research about an
implementation of the Feynman probe mechanism based on current quantum
technology. This would allow us also to extend our analysis
addressing the robustness of the Feynman probe estimation procedure
against experimental imperfections. 
\begin{acknowledgments}
This work has been supported by the EU through the
Collaborative Project QuProCS (Grant Agreement 641277) and by UniMI
through the H2020 Transition Grant 15-6-3008000-625. MGAP thanks K. Zyczkowski for 
useful suggestions.
\end{acknowledgments}
%

\end{document}